\newcommand {\thw}        {\theta_{\mathrm{W}}}
 
\newcommand {\MZ}      {M_{\rm{Z}}}
\newcommand {\MW}      {M_{\mathrm{W}}}
\newcommand {\MH}      {m_{\rm{H}}}
\newcommand {\Mt}      {m_{\rm{t}}}

\newcommand {\GW}      {\Gamma_{\mathrm{W}}}
\newcommand {\Ghad}    {\Gamma_{\mathrm{had}}}
\newcommand {\Gb}      {\Gamma_{\mathrm{b}}}
\newcommand {\Gc}      {\Gamma_{\mathrm{c}}}
\newcommand {\alphamz}    {\alpha(\MZ)}
\newcommand {\alphasmz}    {\alpha_{\rm{s}}{\rm(\MZ)}}
\newcommand {\chisq}       {\chi^{2}}

\newcommand  {\Zzero}   {\mbox{${\mathrm{Z}}$}}
\newcommand {\ff}         {{\rm f}\overline{\rm f}}
\newcommand {\gVf}         {g_{\rm{Vf}}}
\newcommand {\gAf}         {g_{\rm{Af}}}

\newcommand {\gAl}         {g_{\rm{Al}}}
\newcommand {\gVe}         {g_{\rm{Ve}}}
\newcommand {\gAe}         {g_{\rm{Ae}}}
\newcommand {\gVm}         {g_{\rm{V}\mu}}

\newcommand {\gVt}         {g_{\rm{V}\tau}}

\newcommand {\ALR}         {A_{\rm {LR}}}

\newcommand {\swsqa}       {\sin^2\!\thw}

\newcommand {\swsqeffff}   {\sin^2\!\theta_{\rm{eff}}^{\rm {f}}}
\newcommand {\swsqeffl}    {\sin^2\!\theta_{\rm{eff}}^{\rm {lept}}}
\newcommand {\ee}         {\mathrm{e}^+\mathrm{e}^-}
\newcommand {\ppb}        {p\bar{p}}
\newcommand {\roots}      {\sqrt{s}}
\newcommand {\Rbb}        {{R_{\mathrm{b}}}}
\newcommand {\Rcc}        {{R_{\mathrm{c}}}}
\newcommand {\Afbzb}     {A^{0,\,{\rm b}}_{\rm {FB}}}
\newcommand {\Afbzc}     {A^{0,\,{\rm c}}_{\rm {FB}}}
\newcommand {\Ab}      {\rm{A_b}}
\newcommand {\Ac}      {\rm{A_c}}
\newcommand {\Ae}      {\rm{A_e}}
\newcommand {\eeww}       {\ee\rightarrow\mathrm{W^+W^-}}

\newcommand {\qqbp}       {q\bar{q}^{'}}

\documentstyle[12pt,epsfig]{ioplppt}    % PR mod: use this for preprint style

\begin{document}
\title{Electroweak Fits and Constraints on the Higgs Mass }
\author{Peter B. Renton}
\address{\ Denys Wilkinson Building, Keble Road, Oxford OX1 3RH 

e-mail p.renton1@physics.ox.ac.uk}

\vspace*{0.6cm}\begin{center}
Invited contribution to ICHEP 2004, Beijing, China, 16-22 August 2004.
\end{center}

\begin{abstract}
The current status of the quantities entering into the global electroweak fits
is reviewed, highlighting changes since Summer 2003. 
These data include the precision electroweak properties of the Z 
and W bosons, the top-quark mass and the value of the electromagnetic coupling 
constant $\alphamz$, at a scale $\MZ$. 
Using these Z and W (high Q$^{2}$) data, the value of the Higss mass
is extracted, within the context of the Standard Model (SM). 
The consistency of the data,
and the overall agreement with the SM, are discussed.

\end{abstract}

%
%  Uncomment out if preprint format required
%
%\pacs{00.00, 20.00, 42.10}
%\maketitle

\section{The precision electroweak data}%1
This report contains an update on the values of the precision electroweak
properties and fits within the context of the SM, with respect 
to~\cite{ewwg2003}, where more details can be found. 
The $\ee$ data are from the ALEPH, DELPHI, L3 and OPAL experiments at LEP, 
from both the LEP1 and LEP2 phases, and
also from the SLD experiment at SLAC. The $\ppb$ data come from the CDF and
D0 experiments from both Run 1 ($\roots$=1.8 TeV) and Run 2 ($\roots$=1.96 TeV).

\subsection{Z boson}\label{subsec:zzero}%1.1

The coupling of the  $\Zzero$ boson to $\ff$ is specified by the 
vector ($\gVf$) and axial-vector ($\gAf$) couplings. These can be
expressed in terms of $\rho$ and the effective weak mixing angle $\swsqeffff$ by
\begin{equation}
\gAf = \sqrt\rho T_{3}^{f}, \hspace*{0.6cm} 
\gVf/\gAf = 1 - 4\mid q_{f} \mid \swsqeffff 
\end{equation}
where q$_{f}$ is the charge, T$_{3}^{f}$ is the third component 
of weak isospin.
The Z partial width $\Gamma_{f} \propto \gVf^{2} + \gAf^{2}$, and the pole
forward-backward asymmetry, which has been measured 
for e, $\mu$ and $\tau$ pair final states, and also for c and b quarks, is
\begin{equation}
A_{FB}^{0,f} = \frac{3}{4} A_{e}A_{f} , \\
\end{equation}
where
\begin{equation}
A_{f} = \frac {2\gVf/\gAf}{1 + (\gVf/\gAf)^{2}} . 
\end{equation}

The lepton couplings can be extracted from the $\tau$ polarisation 
(giving A$_{e}$, A$_{\tau}$), the SLAC polarised electron 
asymmetry $\ALR$ (A$_{e}$) and the forward-backward asymmetries for 
leptons (A$_{\ell}$, $\ell$=e,$\mu,\tau$). The results are unchanged
with respect to \cite{ewwg2003} and
are reasonably compatible with lepton universality, with
$\gAl$/$\gAe$ = 1.0002 $\pm$ 0.0014 and 1.0019 $\pm$ 0.0015,
for l=$\mu,\tau$ respectively. The uncertainties are larger for the 
vector-couplings, with
$\gVm$/$\gVe$ = 0.962 $\pm$ 0.063 and $\gVt$/$\gVe$ = 0.958 $\pm$ 0.029.
%The $\tau$/e ratios are 1.3 and 1.4 $\sigma$ from unity, for the 
%axial-vector and vector couplings respectively. 
Assuming lepton universality, these asymmetries
give a value of $\Ae$ = 0.1501 $\pm$ 0.0016.
Within the context of the SM this favours a light Higgs mass.
The invisible width of the Z boson allows the number of light neutrinos 
to be extracted (assuming $\Gamma_\nu/\Gamma_l$ from the SM), 
and gives N$_{\nu}$ = 2.9841 $\pm$ 0.0083, which is 1.9 $\sigma$ below 3.
%the SM value of 3.

In the heavy-quark sector there are updates in the results from
SLD. All the LEP and SLD results are now final, but the combination is not yet
finalised. The quantities measured are $\Rbb = \Gb/\Ghad$, $\Rcc = \Gc/\Ghad$,
$\Afbzb$, $\Afbzc$, $\Ab$ and $\Ac$ (which are obtained from
the left-right-forward-backward asymmetries).
There are additional (since Summer 2003) theoretical 
uncertainties, arising from the extrapolation of off-peak measurements 
to the peak, of 0.0002 and 0.0005 added to $\Afbzc$ and $\Afbzb$
respectively (see \cite{rh} for more details). 
There is good internal consistency in the determinations of $\Rbb$, $\Rcc$,
$\Afbzb$ and $\Afbzc$.
The combined LEP and SLD results are given in Table~\ref{tab:s04hf}. The 
largest correlation is -0.18, between $\Rbb$ and $\Rcc$. The $\chi^{2}$/df
for the combination is 53/(105-14), giving a probability close to 100$\%$.
If statistical errors only are used in the combination then this becomes
92/(105-14), indicating that the systematic errors appear to be overestimated.

\begin{table}%1
\caption{Combination of Z heavy flavour results}\label{tab:s04hf}
\begin{center}
\begin{tabular}{|c|c|c|} 
 
\hline 
 
\raisebox{0pt}[12pt][6pt]{quantity} & 
 
\raisebox{0pt}[12pt][6pt]{value} & 
 
\raisebox{0pt}[12pt][6pt]{error} \\
 
\hline

\raisebox{0pt}[12pt][6pt]{$\Rbb$ } & 

\raisebox{0pt}[12pt][6pt]{0.21630} & 
 
\raisebox{0pt}[12pt][6pt]{0.00066} \\

\raisebox{0pt}[12pt][6pt]{$\Rcc$ } & 

\raisebox{0pt}[12pt][6pt]{0.1723} & 
 
\raisebox{0pt}[12pt][6pt]{0.0031} \\

\raisebox{0pt}[12pt][6pt]{$\Afbzb$ } & 

\raisebox{0pt}[12pt][6pt]{0.0998} & 
 
\raisebox{0pt}[12pt][6pt]{0.0017} \\

\raisebox{0pt}[12pt][6pt]{$\Afbzc$ } & 

\raisebox{0pt}[12pt][6pt]{0.0706} & 
 
\raisebox{0pt}[12pt][6pt]{0.0035} \\

\raisebox{0pt}[12pt][6pt]{$\Ab$ } & 

\raisebox{0pt}[12pt][6pt]{0.923} & 
 
\raisebox{0pt}[12pt][6pt]{0.020} \\

\raisebox{0pt}[12pt][6pt]{$\Ac$ } & 

\raisebox{0pt}[12pt][6pt]{0.670} & 
 
\raisebox{0pt}[12pt][6pt]{0.027} \\

\hline
\end{tabular}
\end{center}
\end{table}

%----------------------------------------------------------
\begin{figure}[t]
\vspace*{13pt}
\begin{center}
\mbox{
\epsfig{file=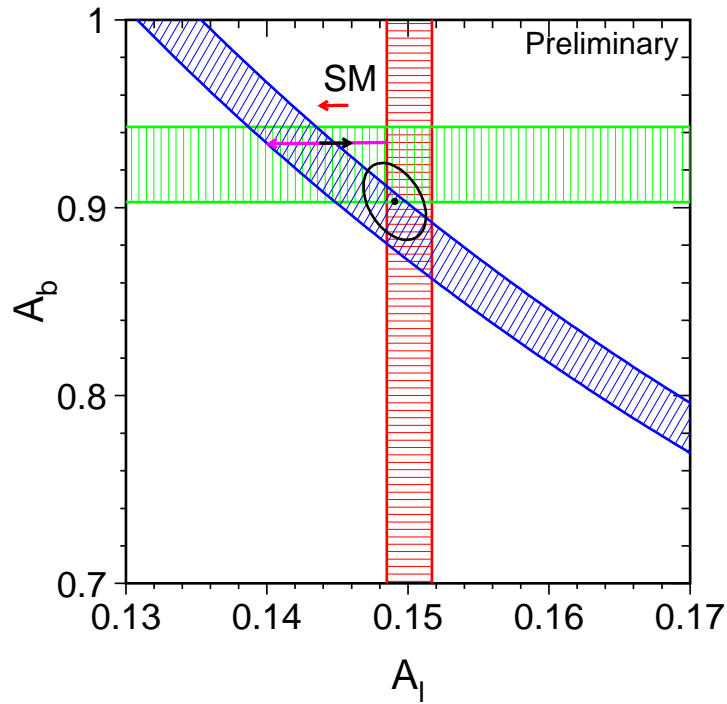,height=11cm}}
\end{center}

\caption{The couplings $\Ab$ and $\Ae$, both from direct measurements and
from $\Afbzb$.}
\label{fig:abvae}
\end{figure}

The direct determinations of $\Ae$ and $\Ab$ are shown in 
figure~\ref{fig:abvae}. Also shown is the band in the $\Ae$ $\Ab$ plane,
traced out by $\Afbzb$. The combined value, and the 68$\%$ cl, are also shown,
as is the SM prediction. It can be seen that the joint result from these data
is in poor agreement with the SM. The value of  $\Afbzb$ favours a 
rather heavy Higgs mass. 

\begin{figure}[t]
\vspace*{13pt}
\begin{center}
\mbox{
\epsfig{file=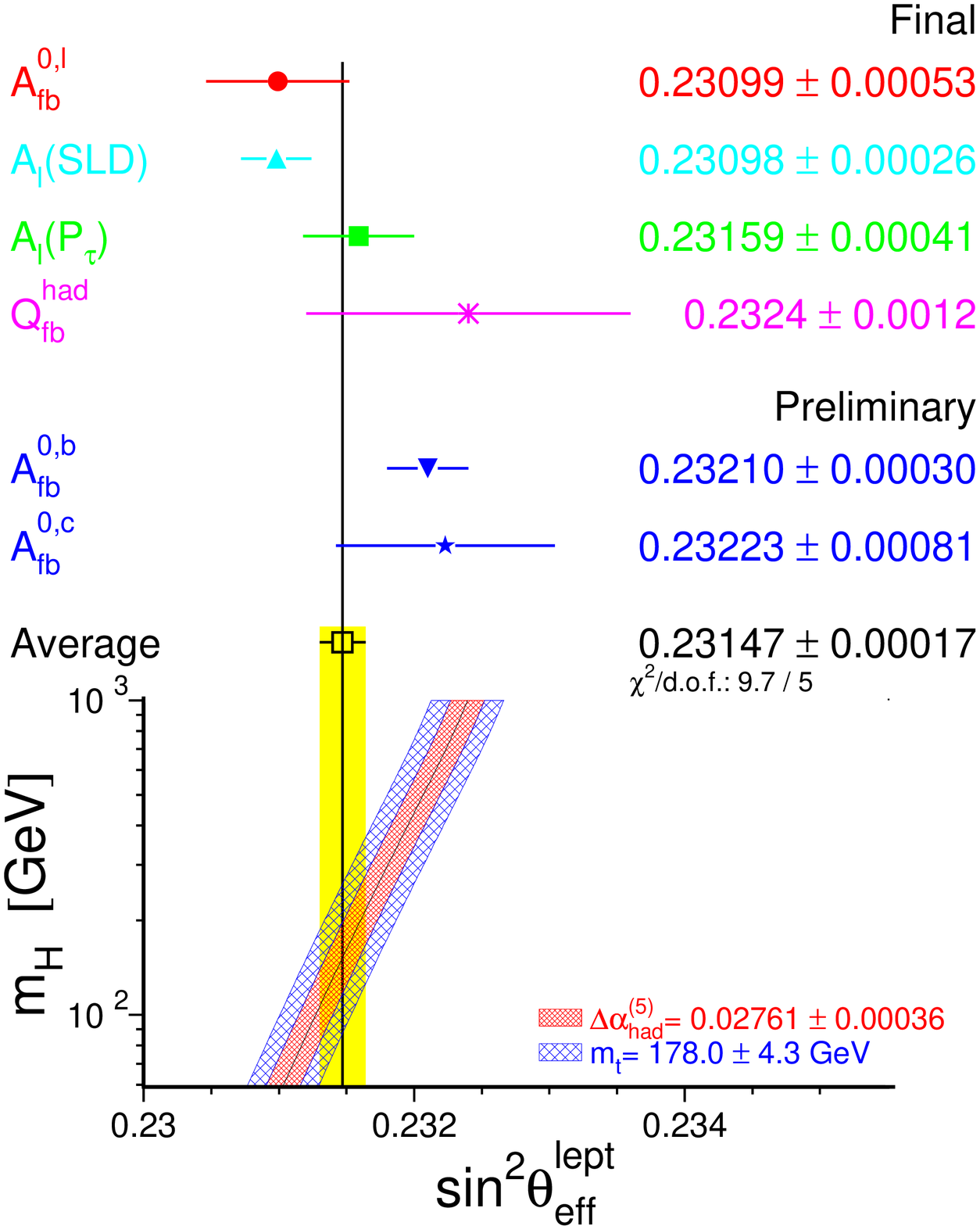,height=11cm}}
\end{center}
\caption{Determinations of $\swsqeffl$.}
\label{fig:s2eff}
\end{figure}

Figure~\ref{fig:s2eff} shows the determinations of $\swsqeffl$. 
The overall $\chisq$ probability is reasonable (8.4$\%$), 
but the value obtained
from purely leptonic processes ($\swsqeffl$=0.23113 $\pm$ 0.00021) is some
2.8$\sigma$ different to that obtained using heavy quarks 
($\swsqeffl$=0.23213 $\pm$ 0.00029). This comes mostly
from the 2.8$\sigma$ difference in the SLD $\ALR$ and $\Afbzb$ values.

%----------------------------------------------------------

\subsection{W boson}\label{subsec:wboson}%1.2

The W boson is produced singly at the Tevatron (eg 
$u+\bar{d} \rightarrow W^{+}$). The leptonic decays W$\rightarrow \ell\nu$ 
(with $\ell = e,\mu$) are used to determine the W mass and width, using 
the transverse mass or p$_{T}^{\ell}$. From Run 1 
the values $\MW$ = 80.433 $\pm$ 0.079 GeV (CDF) and 80.483 $\pm$ 0.084 GeV (D0)
were obtained. Taking into account common systematics, the combined Run 1
values are  $\MW$ = 80.452 $\pm$ 0.059 GeV 
and $\GW$ = 2.102 $\pm$ 0.106 GeV~\cite{tevewwg}.
Run 2 analyses are currently underway.

At LEP2 the W bosons are pair-produced in $\eeww$. The analyses are still 
in progress. The statistical uncertainties from the
$\ell\nu\qqbp$ and $\qqbp\qqbp$ channels are similar. However, there is
at present a large systematic uncertainty (97 MeV) in the $\qqbp\qqbp$ channel,
due to final-state interaction effects. This is mostly from colour reconnection,
with a smaller contribution from Bose Einstein correlations. This means that 
the  $\qqbp\qqbp$ channel carries
only 10$\%$ of the weight in the LEP2 average. The preliminary LEP2 values
are  $\MW$ = 80.412 $\pm$ 0.042 GeV and $\GW$ = 2.152 $\pm$ 0.091 GeV.

The combined Tevatron and LEP2 values are
$\MW$ = 80.425 $\pm$ 0.034 GeV and $\GW$ = 2.133 $\pm$ 0.069 GeV.
$\GW$ is compatible with the SM value of 2.097 $\pm$ 0.003 GeV.
The world average $\MW$ value favours a low Higgs mass in the context of the SM. 

\section{The SM parameters}%2

The SM parameters are taken to be $\MZ$, G$_{F}$,
$\alphamz$ and $\alphasmz$ (the electromagnetic and strong coupling constants
at the scale $\MZ$), and the top-quark mass $\Mt$. Through loop diagrams 
measurements
of the precision electroweak quantities are sensitive to $\Mt$ and, the `unknown'
in the SM, $\MH$. The SM computations use the programs TOPAZ0 and ZFITTER. The
latter program (version 6.40) incorporates the recent fermion 2-loop corrections
to $\swsqeffl$ and full 2-loop, and leading 3-loop, 
corrections to $\MW$~\cite{loops}.

\subsection{top-quark mass}\label{subsec:tquark}%2.1

The D0 Collaboration have recently improved their Run 1 measurement using
a weighting method based on the matrix element, 
giving $\Mt$ = 179.0 $\pm$ 3.5 (stat) $\pm$ 3.8 (syst) GeV. The CDF Run 1 
value is $\Mt$ = 176.1 $\pm$ 4.2 (stat) $\pm$ 5.1 (syst) GeV. Taking into account
common systematic uncertainties the combined value is~\cite{mt}
$\Mt$ = 178.0 $\pm$ 4.3 GeV, with statistical and systematic error components
of 2.7 and 3.3 GeV respectively. This is to be compared to the previous value
of $\Mt$ = 174.3 $\pm$ 5.1 GeV. 

Run 2 values have been obtained by both the CDF and D0 Collaborations, but
these have not yet been included in the average.

\subsection{$\alphamz$}\label{subsec:alpha}%2.2

The value of $\alpha$ at the scale $\MZ$ requires the use of data on
$\ee\rightarrow$hadrons at low energies and the use of perturbative QCD
at higher energies. The various estimations of $\alphamz$ differ 
%in the scale from which QCD is used, 
in the extent to which QCD is used,
as well as in the data used in the evaluation. The
quantity needed is the hadronic contribution $\Delta\alpha^{(5)}_{\rm{had}}$
and the value used by the LEP EWWG~\cite{ewwg2003} is 
$\Delta\alpha^{(5)}_{\rm{had}}$($\MZ$) = 0.02761 $\pm$ 0.00036. Recent data
from the CMD-2 and KLOE Collaborations has been consider in~\cite{alpha},
and the authors conclude that the value just quoted is still valid.

\section{Electroweak fits}%3

\begin{figure}[t]
\vspace*{13pt}
\begin{center}
\mbox{
\epsfig{file=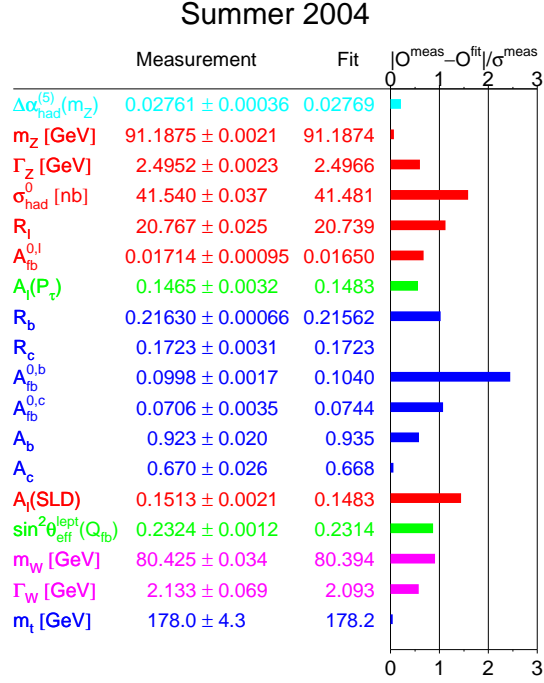,height=11cm}}
\end{center}
\caption{Measured and SM fitted values of electroweak quantities.}
\label{fig:pulls}
\end{figure}

The measurements used in the global SM electroweak fits, and the fitted values, 
are shown in figure~\ref{fig:pulls}. The SM fit to these high Q$^{2}$ data
gives

\begin{center}
$m_{t}$  =  178.2 $\pm$ 3.9  GeV

$m_{H}$  =  114 $^{ +69}_{ -45}$  GeV 

$\alpha_{s}(\MZ)$  =  0.1186 $\pm$ 0.0027. 
\end{center}

%----------------------------------------------------------

\begin{figure}[t]
\vspace*{13pt}
\begin{center}
\mbox{
\epsfig{file=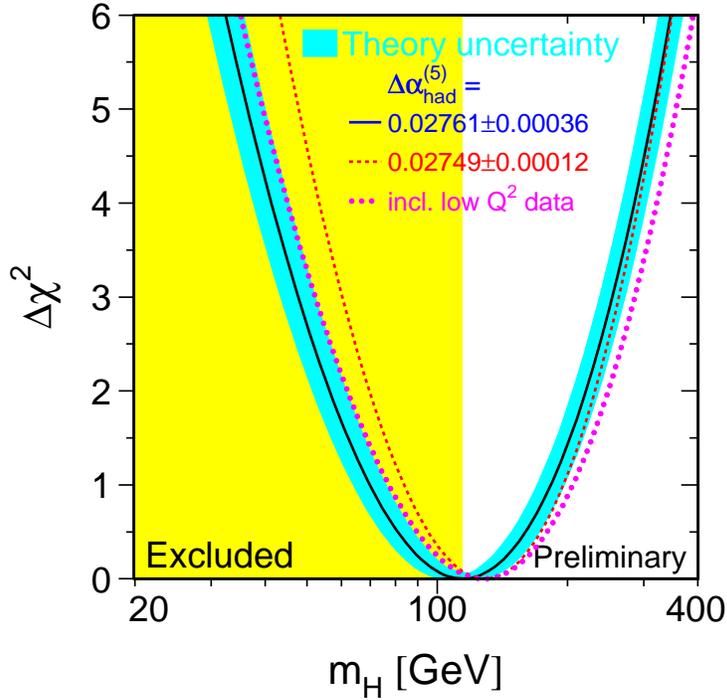,height=11cm}}
\end{center}
\caption{Variation of $\chisq$ versus $\MH$.}
\label{fig:blue}
\end{figure}

The $\chisq$/df is 15.8/13, giving a probability of 26$\%$.
 The variation of the fit $\chisq$, compared to the
minimum value, is shown in the `blue-band' plot of figure~\ref{fig:blue},
as a function of $\MH$. Also shown is the direct search limit of 114 GeV.
The one-sided 95$\%$ upper limit is $\MH \leq$ 260 GeV. This includes the
theoretical uncertainty (blue-band) which is evaluated by considering 
the uncertainties in the new 2-loop calculations~\cite{loops}.
If the more theory driven value 
$\Delta\alpha^{(5)}_{\rm{had}}$($\MZ$) = 0.02749 $\pm$ 0.00012 is used, then
$\MH$ increases to 129 GeV.

%----------------------------------------------------------

Since 2003 the main changes have been the change in $\Mt$ 
($\delta\MH \simeq$ +20 GeV) and the new 2-loop 
effects ($\delta\MH \simeq$ +6 GeV).

\begin{figure}[t]
\vspace*{13pt}
\begin{center}
\mbox{
\epsfig{file=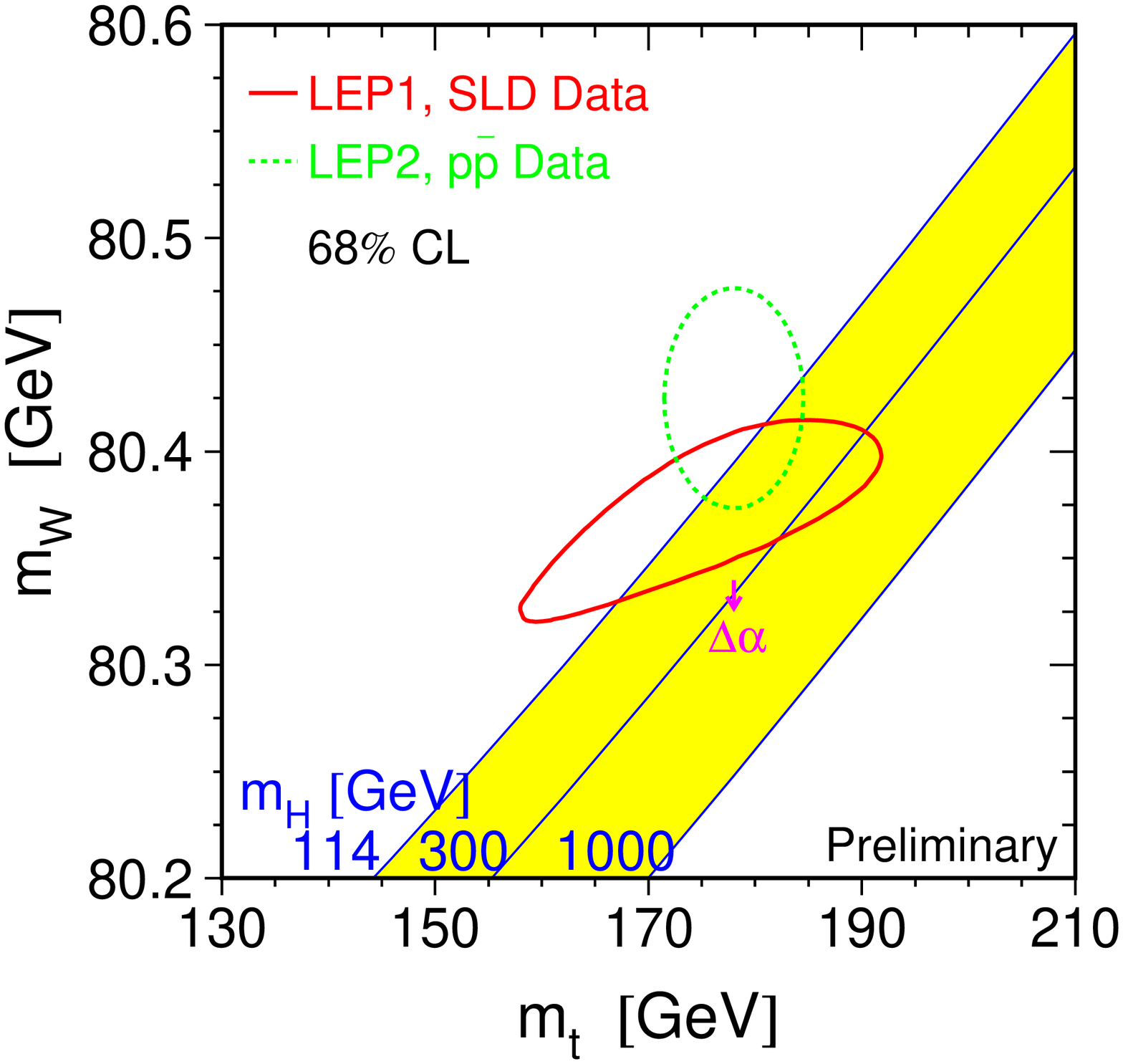,height=11cm}}
\end{center}
\caption{Direct versus indirect $\Mt$ and $\MW$ measurements.}
\label{fig:dirvindir}
\end{figure}

The direct versus indirect values of $\Mt$ and $\MW$ is a powerful test
of the SM; see figure~\ref{fig:dirvindir}. The contours shown are for the
68$\%$ cl. It can be seen that there is a reasonable degree of overlap 
and that the data prefer a light Higgs mass.

%----------------------------------------------------------

The above fits use only high Q$^{2}$ data. There are also 
low Q$^{2}$ data\cite{pl} from Atomic Parity Violation in $^{133}$Cs 
(Q$_{W}$ = -72.74 $\pm$ 0.46), the SLAC polarised electron Moller scattering
experiment E158 ($\swsqeffl$ = 0.2333 $\pm$ 0.0016) and the deep-inelastic
$\nu(\bar{\nu})$ experiment NuTeV ($\swsqa$ = 0.2277 $\pm$ 0.0016). The NuTeV
value can be used to extract $\MW$, and gives a value 3.1$\sigma$ below that
from direct measurement. Including all these low Q$^{2}$ data in the SM fit
increases $\MH$ by 14 GeV to 128 GeV, and the $\chisq$ probability drops
to 5.4$\%$, essentially due to the NuTeV result. 

\section{Conclusions}%4
There has been steady progress on both the experimental and theoretical fronts.
There are still issues with $\Afbzb$ and NuTeV (both $\simeq$3$\sigma$ 
effects).
%, but it is difficult to see how these can be fully resolved in the near future. 
It is difficult to see how $\Afbzb$ can be resolved in the near future, but
for NuTeV, the further evaluation of QED and QCD effects, together with
the NOMAD results, should help.

The SM fits favour a light Higgs mass, 
$\MH$ = 114 $^{ +69}_{ -45}$ GeV, and a 95$\%$ cl upper limit of 260 GeV.
Thus the Higgs boson appears to be relatively light. Improved measurements
of both $\Mt$ and $\MW$ at the Tevatron, and then the LHC, will significantly
improve the precision of the indirect estimation of $\MH$.

\section*{Acknowledgments}
I wish to thank the LEP EWWG, in particular Martin Gr\"{u}newald, for their help.

\end{document}